# Syndrome-Enabled Unsupervised Learning for Neural Network-Based Polar Decoder and Jointly Optimized Blind Equalizer

Chieh-Fang Teng, *Student Member, IEEE*, and Yen-Liang Chen

*Abstract*—Recently, the syndrome loss has been proposed to achieve "unsupervised learning" for neural network-based BCH/LDPC decoders. However, the design approach cannot be applied to polar codes directly and has not been evaluated under varying channels. In this work, we propose two modified syndrome losses to facilitate unsupervised learning in the receiver. Then, we first apply it to a neural network-based belief propagation (BP) polar decoder. With the aid of CRC-enabled syndrome loss, the BP decoder can even outperform conventional supervised learning methods in terms of block error rate. Secondly, we propose a jointly optimized syndrome-enabled blind equalizer, which can avoid the transmission of training sequences and achieve global optimum with 1.3 dB gain over non-blind minimum mean square error (MMSE) equalizer.

## I. Introduction

WITH more and more revolutionized breakthroughs in the field of computer vision and natural language processing, machine learning-assisted communication systems have also attracted a lot of researchers in this newly emerging field. Most of the well-designed neural networks are either as the replacements for certain blocks [1]–[6] or as an end-to-end solutions [7]–[10]. In [1]–[2], convolutional neural networks (CNNs) are exploited for powerful modulation classification. For channel decoding, neural network-based decoders are proposed in [3]–[5], which assign trainable weights to the factor graph and thus improve the convergence speed with better performance. In [6], a neural network-based equalizer is proposed to utilize coding gain in advance. Furthermore, for end-to-end optimization, by replacing the whole system with neural networks, the authors in [7]–[10] try to break the conventional rules of independent block design by jointly optimizing the whole communication systems.

Though neural networks can achieve promising performance in the simulation experiments, one of the unique and realistic challenges is that the physical channel or hardware impairments will vary over time and temperature variation. However, most of the "trainable" communication systems are based on supervised learning, which neglects the feasibility and overhead of obtaining labeled training data in practical applications. Without the precious training data for accurate channel estimation, the performance degrades severely under time-varying channels or fluctuations of hardware impairments [11]–[14]. Therefore, unsupervised learning plays an important role to overcome the challenge of online channel adaptation without incremental transmission overhead.

Recently, some approaches for unsupervised learning are proposed [13]–[15]. In [13]–[14], the mechanism of online label recovery is proposed to take advantage of error correction code (ECC) to correct the corrupted signals and the re-encoded signals can be collected as labeled training set. Therefore, it can compensate for slow fluctuations in channel conditions and hardware impairments without any transmission overhead. On the other hand, syndrome loss was proposed in [15]. By penalizing the decoder for producing non-valid codewords, it can be used to train neural network-based decoder for Bose-Chaudhuri-Hocquenghem (BCH) codes and low-density parity-check (LDPC) codes without prior knowledge of the transmitted codewords. As a result, online label recovery and syndrome loss can provide promising tools for online channel adaptation solutions for unsupervised learning. However, there are still some issues need to be addressed:

1) *Constraints of Online Label Recovery:* The mechanism of online label recovery relies on the correction of all erroneous bits in the corrupted signals, so the re-encoded codeword will be completely correct. Otherwise, the incorrectly decoded output will result in incorrectly re-encoded codeword for training and even degrade the system performance. Therefore, this mechanism demands for a high signal-to-noise ratio (SNR) to ensure low error rate.

2) *Constraints of Syndrome Loss:* This approach demands the decoder to output the soft estimation of the codewords [15]. Thus, it can be examined by parity-check

Manuscript received December 29, 2019; revised March 17, 2020 and April 16, 2020; accepted April 30, 2020. Date of publication May 6, 2020; date of current version June 12, 2020. This work was supported by MediaTek, Inc., Hsinchu, Taiwan, under Grant MTKC-2019-0070. Chieh-Fang Teng was also supported by the MediaTek Ph.D. Fellowship Program. This article was recommended by Guest Editor Y.-L. Ueng. *(Corresponding author: Chieh-Fang Teng.)*

Chieh-Fang Teng is with the Graduate Institute of Electronics Engineering (GIEE), National Taiwan University, Taipei 10617, Taiwan (e-mail: jeff@access.ee.ntu.edu.tw).

Yen-Liang Chen is with MediaTek, Inc., Hsinchu 30078, Taiwan (e-mail: ben@access.ee.ntu.edu.tw).

Color versions of one or more of the figures in this article are available online at https://ieeexplore.ieee.org.

Digital Object Identifier 10.1109/JETCAS.2020.2992593

matrix to produce syndrome loss. Unfortunately, polar decoders [16], whose outputs are source bits without the definition of parity-check matrix, cannot directly apply syndrome loss for unsupervised learning.

3) *Inadequate Application Scenario:* In [15], the authors evaluate the capability of proposed syndrome loss only on neural network-based decoder under additive white Gaussian noise (AWGN) channel. In such scenario, the labeled data can be obtained easily by conducting simulations in AWGN channels. Therefore, this application scenario cannot really reveal the benefits and prominences of syndrome loss.

In this paper, we aim to exploit more possibilities and extend the potential usages of syndrome loss. In order to address the aforementioned issues of syndrome loss, we propose two kinds of modified syndrome losses to enable unsupervised learning for polar codes. Then, two application scenarios are also proposed to evaluate and demonstrate their capabilities. The main contributions of this paper are summarized as follows:

1) *Define Two Kinds of Modified Syndrome Losses:* By exploiting the nature of polar codes and taking advantage of the standardized cyclic redundancy check (CRC) mechanism in the fifth generation (5G) communication systems [17], we propose two kinds of modified syndrome losses, frozen-bit syndrome loss and CRC-enabled syndrome loss, which enable unsupervised learning to be used for polar codes.

2) *Apply Unsupervised Learning to Neural Network-Based Belief Propagation Polar Decoder:* We firstly evaluate and compare the performance between the widely adopted supervised learning and unsupervised learning for neural network-based decoder, which use binary cross-entropy (BCE) and the proposed two kinds of syndrome losses, respectively. From the simulation results, the proposed CRC-enabled syndrome loss can even outperform supervised learning in terms of block error rate (BLER) due to its block-level optimization.

3) *Propose Syndrome-Enabled Blind Equalizer With Joint Optimization Mechanism:* We consider the block fading channel in this part, which means it demands training sequence for channel adaptation. The prior frameworks are shown in Fig. 1(a) and Fig. 1(b), and the proposed syndrome-enabled blind equalizer with joint optimization is shown in Fig. 1(c). From the simulation results, the proposed method can improve spectral efficiency by avoiding the transmission of training sequence and achieve global optimum via joint optimization mechanism with 1.3 dB gain. The feature comparison of different frameworks is summarized in Table I.

The remaining part of this paper is organized as follows. In Section II, the system models, prior works of channel equalization, online label recovery, polar codes, and syndrome loss are introduced. Section III derives the proposed frozen-bit syndrome loss and CRC-enabled syndrome loss. In Section IV, the application scenario of unsupervised learning for neural network based decoder is proposed with conducted simulation results for evaluation. In Section V, we propose another application scenario of syndrome-enabled blind equalizer with complete simulation results and detailed analysis. The conclusions are finally drawn in Section VI.

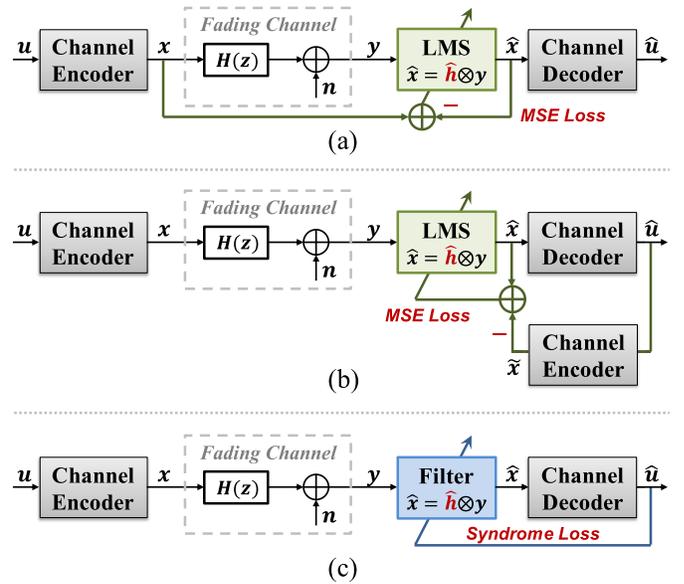

Fig. 1. Overview of system model with different approaches for updating of filter coefficients: (a) conventional equalizer with least mean squares (LMS) algorithm [18], (b) blind equalizer with the mechanism of online label recovery, [13], and (c) the proposed syndrome-enabled blind equalizer with the mechanism of joint optimization.

TABLE I
FEATURE COMPARISON OF DIFFERENT FRAMEWORKS
FOR UPDATING OF FILTER COEFFICIENTS

|  | Adaptive MMSE [18] | MMSE with Online Label Recovery [13] | Proposed Syndrome-Enabled Blind Equalizer |
|---|---|---|---|
| Requirement of Training Sequence | YES | NO | NO |
| Requirement of Decoding | NO | YES | YES |
| Requirement of High SNR | NO | YES | NO |
| Joint Optimization Mechanism | NO | NO | YES |

## II. SYSTEM MODEL AND BACKGROUND

### A. Notations and System Model

Throughout this paper, a normal-faced letter $a$ denotes a scalar, a bold-faced lowercase letter $\boldsymbol{a}$ denotes a vector, and a bold-faced uppercase letter $\mathbf{A}$ denotes a matrix. Other operations used in this paper are defined as follows:

- $a_p$ represents the $p$-th element of $\boldsymbol{a}$.
- $\mathbf{A}_{p,:}$, $\mathbf{A}_{:,p}$, and $\mathbf{A}_{p,q}$ denote the $p$-th row vector, $q$-th column vector, and $(p, q)$-th entry of $\mathbf{A}$, respectively.
- $\mathbf{A}^T$ denotes the transpose of $\mathbf{A}$.
- $|\cdot|$ denotes the element-wise absolute value or cardinality for a set.

The overall system model is depicted in Fig. 1. At the transmitter side, the information messages $u$ are first encoded as $x$ by the encoder and then sent to the channel. Throughout this paper, a block fading channel is considered. Therefore, inter-symbol interference (ISI) and additive white Gaussian noise (AWGN) jointly contribute to channel distortion and the received signal can be expressed as:

$$y_i = \sum_{l=0}^{L-1} x_{i-l} \times h_l + n_i, i = 0, \ldots, N-1, \quad (1)$$

where $L$ is the length of the response, $h$ and $n$ are the channel impulse response and the additive white Gaussian noise (AWGN) with variance $\sigma^2$, respectively, and $N$ denotes the codeword length.

At the receiver side, the equalizer is firstly applied to eliminate the ISI effects, which will be reviewed in Section II.B. Then, based on the reconstructed signal $\hat{x}$, output from the equalizer, a decoder is used to do error correction and estimate $\hat{u}$ for the original information messages $u$.

### B. Prior Works of Channel Equalization [18]–[25]

Inter-symbol interference occurs when the transmitted signal has multiple paths to reach the receiver, which results in combination of symbols over time with severe interference. The process of removing the ISI is called equalization and minimum mean square error (MMSE) through least mean squares (LMS) algorithm is one of the widely adopted adaptive filters for channel equalization [18]. The system with MMSE equalizer is shown in Fig. 1(a). The main idea of MMSE equalizer is to adapt a filter $\hat{h}$ such that the convolution with received signal $y$ is close to the training sequences $d$ by minimizing the MMSE cost function and can be expressed as:

$$e_i = E\left\{\left|d_i - \sum_{l=0}^{F-1} y_{i-l} \times \hat{h}_l\right|^2\right\}, \quad (2)$$

where $F$ is the number of filter coefficients and $e$ is the minimized target by gradient descent algorithm to obtain optimum filter weights.

As the channel will vary over time, adaptive MMSE equalizers require training sequences for channel adaptation, which degrades the spectral efficiency. On the other hand, blind equalizers recover the signal merely based on the observed channel outputs and the known statistics of channel input signal, which do not require training sequences [19]–[25]. For example, the constant modulus algorithm (CMA) [19]–[21], forces the output of the equalizer to have constant amplitude with the following cost function:

$$J_{CMA,i} = E\left\{\left[|\hat{x}_i|^2 - R_2\right]^2\right\}, R_2 = E\left\{|x_i|^4\right\}/E\left\{|x_i|^2\right\}, \quad (3)$$

where $R_2$ is Godard dispersion constant depending on a priori statistical information about transmitted signal. Then, the coefficients of filter can also be updated by minimizing the cost function via gradient descent.

In summary, although blind equalizers can avoid the transmission of training sequences and thus improve spectral efficiency, the quality of reconstructed signal is not as well as non-blind equalizers with training sequences.

### C. Online Label Recovery [13]–[14]

As more and more researchers focus on the field of machine learning assisted communication systems, one of the unique challenges in this field is that the physical channel or hardware impairments will vary over time and temperature variation. Therefore, how to acquire labeled data for finetuning of trainable communication systems without sacrificing spectral efficiency is a severe issue that needs to be addressed.

In [13] and [14], the authors propose a new concept to recover a labeled training set on the receiver side without any transmission overhead of pilot signals as shown in Fig. 1(b). Both of them utilize error correction code (ECC) to correct the corrupted signals. Suppose all error bits in the equalized codeword $\hat{x}$ can be completely corrected, the decoded message bits $\hat{u}$ will be re-encoded to provide the codeword $\tilde{x}$. Therefore, a training set can be constructed during data transmission and utilized to finetune model parameters via stochastic gradient descent (SGD) algorithm as below:

$$\theta_{new} = \arg\min_{\theta} \mathcal{L}\left(\hat{x}, \tilde{x}\right), \quad (4)$$

where $\mathcal{L}$ is the pre-defined loss function and $\theta$ is the model parameters. In our case, $\theta_{new}$ is the updated filter coefficients.

The authors have demonstrated the capability of this technique by finetuning the system on-the-fly to compensate for the slow fluctuations of IQ-imbalance, hardware non-linearity, and channel conditions without any transmission overhead. However, all of the results are based on the premise that ECC can recover all error bits in the codeword, otherwise the incorrectly re-encoded codeword $\tilde{x}$ will even deteriorate the model parameters instead.

### D. Polar Codes and Belief Propagation Decoding Algorithm [26]–[28]

To construct an $(N, K)$ polar code, the $N$-bit message $u$ is recursively constructed from a polarizing matrix $\mathbf{F} = \begin{bmatrix} 1 & 0 \\ 1 & 1 \end{bmatrix}$ by $\log_2 N$ times to exploit channel polarization [16]. As $N \to \infty$, the synthesized channels tend to two extremes: the noisy channels and noiseless channels. Therefore, the $K$ information bits are assigned to the $K$ most reliable bits in $u$ and the remaining $(N-K)$ bits are referred to as frozen bits with the assignment of zeros. Then, the $N$-bit transmitted codeword $c$ can be generated by multiplying $u$ with generator matrix $\mathbf{G}$ as follows:

$$c = \mathbf{G}u = \mathbf{F}^{\otimes n}\mathbf{B}u, n = \log_2 N. \quad (5)$$

$\mathbf{F}^{\otimes n}$ is the $n$-th Kronecker power of $\mathbf{F}$, and $\mathbf{B}$ represents the bit-reversal permutation matrix.

For the polar decoder, belief propagation (BP) is a widely used algorithm, which can achieve high throughput due to its parallelized architecture [26]–[28]. There are two types of log likelihood ratios (LLRs) iteratively updated on the factor graph, namely left-to-right message $\mathbf{R}_{i,j}^{(t)}$ and right-to-left message $\mathbf{L}_{i,j}^{(t)}$, where node $(i, j)$ represents $j$-th node at the $i$-th stage and $t$ indicates the $t$-th iteration as shown in Fig. 2. Before the iterative propagation and updating, the values of LLR are first initialized as:

$$\mathbf{R}_{0,j}^{(1)} = \begin{cases} 0, & if \; j \in A \\ +\infty, & if \; j \in A^c \end{cases}, \quad \mathbf{L}_{n,j}^{(1)} = \ln \frac{P(y_j|x_j = 0)}{P(y_j|x_j = 1)}, \quad (6)$$

where $A \subseteq \{0, 1, \ldots, N - 1\}$ denotes the set of indices for the information bits and $A^c$ is its complement to represent the set of indices for the frozen bits. Then, the iterative decoding procedure with the updating of $\mathbf{R}_{i,j}^{(t)}$ and $\mathbf{L}_{i,j}^{(t)}$ is given by:

$$\begin{cases} \mathbf{L}_{i,j}^{(t)} = g\left(\mathbf{L}_{i+1,j}^{(t-1)}, \mathbf{L}_{i+1,j+N/2^i}^{(t-1)} + \mathbf{R}_{i,j+N/2^i}^{(t)}\right), \\ \mathbf{L}_{i,j+N/2^i}^{(t)} = g\left(\mathbf{R}_{i,j}^{(t)}, \mathbf{L}_{i+1,j}^{(t-1)}\right) + \mathbf{L}_{i+1,j+N/2^i}^{(t-1)}, \\ \mathbf{R}_{i+1,j}^{(t)} = g\left(\mathbf{R}_{i,j}^{(t)}, \mathbf{L}_{i+1,j+N/2^i}^{(t-1)} + \mathbf{R}_{i,j+N/2^i}^{(t)}\right), \\ \mathbf{R}_{i+1,j+N/2^i}^{(t)} = g\left(\mathbf{R}_{i,j}^{(t)}, \mathbf{L}_{i+1,j}^{(t-1)}\right) + \mathbf{R}_{i,j+N/2^i}^{(t)}, \end{cases} \quad (7)$$

and $g(x, y) \approx \text{sign}(x) \text{sign}(y) \min(|x|, |y|)$ is the min-sum approximation. Finally, after $T$ iterations, the estimation of $\hat{u}$ is decided by:

$$\hat{u}_j = \begin{cases} 0, & if \; \mathbf{L}_{0,j}^{(T)} + \mathbf{R}_{0,j}^{(T)} \geq 0, \\ 1, & if \; \mathbf{L}_{0,j}^{(T)} + \mathbf{R}_{0,j}^{(T)} < 0. \end{cases} \quad (8)$$

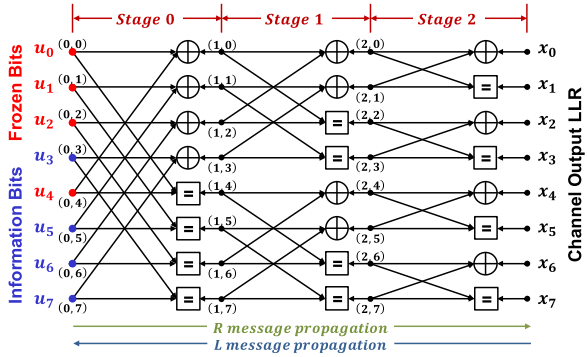

Fig. 2. Factor graph of polar codes with $N = 8$, $A = \{3, 5, 6, 7\}$, and $A^c = \{0, 1, 2, 4\}$.

For more details of the derivations, please refer to the works of [26]–[28].

### E. Syndrome Loss [15]

In [15], the authors introduced syndrome loss, which penalizes the decoder for producing outputs that do not correspond to valid codewords. In communication systems, the transmitter encodes a $K$-bit message $\mathbf{u} \in \text{GF}(2)^K$ by using a generator matrix $\mathbf{G} \in \text{GF}(2)^{N \times K}$ to obtain an $N$-bit codeword $\mathbf{c} = \mathbf{Gu} \in \text{GF}(2)^N$. After transforming to a bipolar format $\mathbf{x} = 1 - 2\mathbf{c} \in \{-1, 1\}^N$, the codeword is transmitted over the channel. Then, the decoder will estimate $\mathbf{x}$ from the received noisy signal $\mathbf{y}$. The estimated bipolar codeword, $\hat{\mathbf{x}} = \text{sign}(\mathbf{s})$, is found by taking the hard decision of soft output from decoder $\mathbf{s} \in \mathbb{R}^N$, and the corresponding estimated binary codeword is $\hat{\mathbf{c}} = 0.5 - 0.5\hat{\mathbf{x}}$.

For a linear code, the estimated binary codeword $\hat{\mathbf{c}}$ can be examined by a parity-check matrix $\mathbf{H} \in \text{GF}(2)^{(N-K) \times N}$, and the syndrome is defined as the product $\mathbf{H}\hat{\mathbf{c}} \in \text{GF}(2)^{N-K}$. For a valid codeword, the syndrome must contain only 0. Therefore, the syndrome can be used to check if the decoder has successfully produced a valid codeword. Based on this concept, a differentiable soft syndrome is defined as follows:

$$\text{softsynd}(\mathbf{s}, \mathbf{H})_i = \min_{j \in \mathcal{M}(i)} |s_j| \prod_{j \in \mathcal{M}(i)} \text{sign}(s_j), \quad (9)$$

where $\mathcal{M}(i)$ is the set of entries in the $i$th row of $\mathbf{H}$ equal to 1 and this equation is extended from the check node update equation in min-sum decoding algorithm.

To maximize each entry in the soft syndrome, the syndrome loss can be constructed to penalize the negative ones, which is defined as:

$$\mathcal{L}_{\text{synd}}(\mathbf{s}, \mathbf{H}) = \frac{1}{N-K} \sum_{i=0}^{N-K-1} \max\left(1 - \text{softsynd}(\mathbf{s}, \mathbf{H})_i, 0\right). \quad (10)$$

Therefore, the loss function can be calculated without the knowledge of transmitted codeword $\mathbf{c}$ and backpropagated for the trainable communication systems under unsupervised learning.

## III. PROPOSED SYNDROME LOSS FOR POLAR CODES

### A. Challenges of Syndrome Loss

For a BCH code or an LDPC code, belief propagation decoding is iteratively performed on a bipartite graph, which is constructed from the well-defined parity-check matrix. Besides, the output of decoder is a soft estimation of codeword and can be directly checked by the matrix, which meets the requirements for using syndrome loss.

However, polar decoders directly estimate the message $\mathbf{u}$ from (8) without providing the definition of $\mathbf{H}$, which restricts the usage of syndrome loss. To address this issue, by exploiting the nature of polar code and taking advantage of the standardized CRC mechanism in 5G, we derive two kinds of modified syndrome losses with suitable parity-check matrices for unsupervised learning.

### B. Frozen-Bit Syndrome Loss

According to aforementioned constraints for using syndrome loss for polar codes, we need to define the parity-check matrix $\mathbf{H}$ with suitable output $\mathbf{s}$ from decoder to produce soft syndrome in (9) for penalization. Though polar code does not provide a parity-check matrix inherently, it has a special characteristic, the frozen bits, which are set to 0 and allow us to derive the parity-check matrix.

Firstly, the codeword $\mathbf{c} = \mathbf{Gu}$ can be inverted as:

$$\mathbf{u} = \mathbf{G}^{-1}\mathbf{c} = \left(\mathbf{F}^{\otimes n}\mathbf{B}\right)^{-1}\mathbf{c} = \left(\mathbf{F}^{-1}\right)^{\otimes n}\mathbf{B}^{-1}\mathbf{c}, \quad (11)$$

where $\mathbf{F}^{-1} = \mathbf{F}$ and $\mathbf{B}^{-1} = \mathbf{B}$. Thus, (11) can be simplified as:

$$\mathbf{u} = \left(\mathbf{F}^{-1}\right)^{\otimes n}\mathbf{B}^{-1}\mathbf{c} = \mathbf{F}^{\otimes n}\mathbf{B}\mathbf{c} = \mathbf{Gc}. \quad (12)$$

Then, based on the characteristic of polar codes that the frozen bits are set to 0, $\boldsymbol{u}$ with indices of the frozen bits must be 0. Thus, (12) can be further derived as:

$$\boldsymbol{u}_{A^c} = \mathbf{G}_{A^c}\boldsymbol{c} = \mathbf{0}. \quad (13)$$

Consequently, we can conclude that the specific parity-check matrix $\mathbf{H}_{\text{froz}}$ for polar code is as follows:

$$\mathbf{H}_{\text{froz}} = \mathbf{G}_{A^c}, \quad (14)$$

which is formed from the rows of $\mathbf{G}$ with indices in $A^c$. The soft output from decoder $\boldsymbol{s}_{\text{froz}}$ can be obtained as:

$$s_{\text{froz},j} = \mathbf{L}_{n,j}^{(T)} + \mathbf{R}_{n,j}^{(T)}, \forall j \in \{0, \ldots, N-1\}, \quad (15)$$

which is used for the calculation of syndrome loss. Besides, we adopt the technique of multi-loss to improve the stability and convergence speed of the training process as proposed in [3]. Therefore, the estimated soft output $\boldsymbol{s}_{\text{froz}}$ for each iteration can be obtained by:

$$s_{\text{froz},j}^t = \mathbf{L}_{n,j}^{(t)} + \mathbf{R}_{n,j}^{(t)}, \forall j \in \{0, \ldots, N-1\} \text{ and } \forall t \in \{1, \ldots, T\}. \quad (16)$$

Therefore, the function of frozen-bit syndrome loss with average loss among all iterations can be defined as follows:

$$\mathcal{L}_{\text{froz\_synd}}(\boldsymbol{s}_{\text{froz}}, \mathbf{H}_{\text{froz}}) = \frac{1}{T}\sum_{t=1}^{T}\left[\frac{1}{N-K}\sum_{i=0}^{N-K-1} \max\left(1 - \text{softsynd}\left(\boldsymbol{s}_{\text{froz}}^t, \mathbf{H}_{\text{froz}}\right)_i, 0\right)\right]. \quad (17)$$

### C. CRC-Enabled Syndrome Loss

Polar codes have drawn a lot of attention due to their provable capacity-achieving [16]. However, the error-correction performance for short to moderate code lengths does not meet the requirements in 5G. Thus, many methods are proposed to enhance the performance. Among the various approaches, cyclic redundancy check (CRC) contributes a vast improvement for SC list (SCL) [29]. Generally, CRC is utilized to check the correctness of decoded results. Thus, it is useful for SCL to select the right path or as an early termination criterion for BP to prevent unnecessary iterations.

Furthermore, the authors of [30] concatenate the CRC factor graph with the polar factor graph as shown in Fig. 3. By running BP decoding on the concatenated factor graph, the authors demonstrate that the CRC-aided BP decoder has significant error-correction performance improvement over the conventional BP decoder from simulation results. Besides, they also assign trainable weights to the edges of the factor graph, which further enhances the performance under limited number of BP iterations.

As the mechanism of CRC is standardized in enhanced mobile broadband (eMBB) control channel of 5G, we can expect that polar codes will equip this mechanism in the future. Inspired from the concatenated factor graph as shown in Fig. 3, though polar code does not provide parity-check matrix, the part of CRC does have a smaller parity-check matrix $\mathbf{H}_{\text{CRC}}$, which allows the utilization of syndrome loss to penalize the codeword $\boldsymbol{x}_{\text{CRC}}$. Besides, though the restrictions are only added to the part of CRC factor graph, the codeword is jointly decided by the whole factor graph. Consequently, the whole polar factor graph can also be optimized via backpropagation.

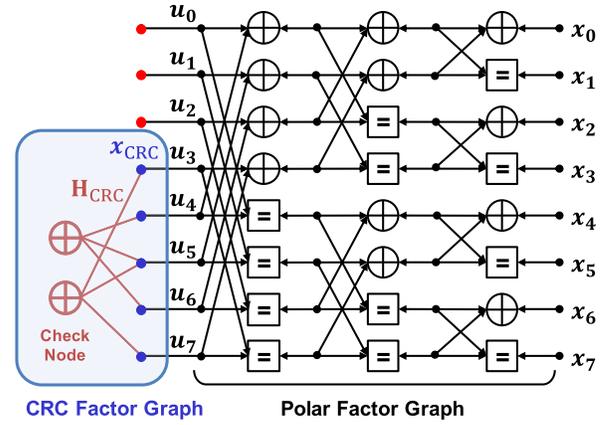

Fig. 3. Factor graph of a CRC-polar concatenated code with $N = 8$, $K = 3$ and 2-bit CRC is used.

In this part, the technique of multi-loss is also adopted. Therefore, the estimated soft output $\boldsymbol{s}_{\text{CRC}}$ for each iteration can be obtained by:

$$s_{\text{CRC},j}^t = \mathbf{L}_{0,j}^{(t)} + \mathbf{R}_{0,j}^{(t)}, \forall j \in A \text{ and } \forall t \in \{1, \ldots, T\}. \quad (18)$$

Similarly, the function of CRC-enabled syndrome loss with average loss among all iterations can be defined as follows:

$$\mathcal{L}_{\text{CRC\_synd}}(\boldsymbol{s}_{\text{CRC}}, \mathbf{H}_{\text{CRC}}) = \frac{1}{T}\sum_{t=1}^{T}\left[\frac{1}{C}\sum_{i=0}^{C-1} \max\left(1 - \text{softsynd}\left(\boldsymbol{s}_{\text{CRC}}^t, \mathbf{H}_{\text{CRC}}\right)_i, 0\right)\right], \quad (19)$$

where $C$ is the number of CRC bits. In Section IV and Section V, we will exemplify two different application scenarios to evaluate the capabilities for unsupervised learning of these two syndrome loss and compare their performances with prior works.

## IV. APPLICATION SCENARIO I: UNSUPERVISED LEARNING FOR NEURAL NETWORK-BASED POLAR DECODER

One of the many advantages of machine learning assisted communication systems is that the trainable communication systems can achieve better performance based on the training data obtained from the current channel state. However, in real-world wireless communications, most of the channel conditions and hardware impairments will vary over time, such as Doppler shifts, inter-symbol interference, and hardware non-linearity, which results in mismatch between training data and inference data. Thus, trainable communication systems demand finetuning to adapt to new state and maintain the advantages over conventional communication systems.

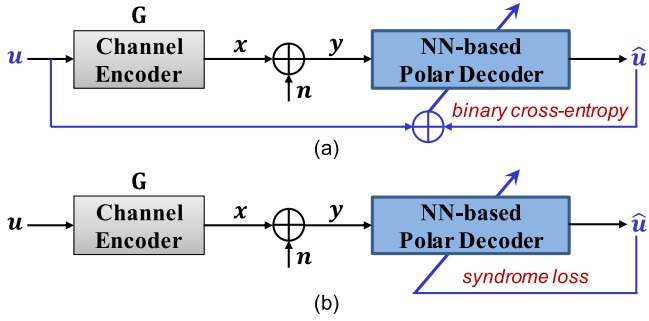

Fig. 4. Training process for neural network-based polar decoder: (a) supervised learning with binary cross-entropy loss; (b) unsupervised learning with syndrome loss.

Based on the two different kinds of proposed syndrome loss, we want to demonstrate that the trainable systems can be finetuned without transmission of training sequence. Therefore, our approach can maintain the high performance without increased transmission overhead.

### A. Proposed Unsupervised Learning for Neural Network-Based Belief Propagation (NN-BP) Polar Decoder

In [4]–[5], by taking advantage of recent advances in deep learning (DL), a neural network-based belief propagation (NN-BP) polar decoder is proposed with assigned weights on the factor graph. After training, the weights function as the scaling factor for the importance of messages and the iterative decoding procedure derived from (7) is revised as:

$$\begin{cases} \mathbf{L}_{i,j}^{(t)} = \alpha_{i,j}^{(t)} \cdot g\left(\mathbf{L}_{i+1,j}^{(t-1)}, \mathbf{L}_{i+1,j+N/2^i}^{(t-1)} + \mathbf{R}_{i,j+N/2^i}^{(t)}\right), \\ \mathbf{L}_{i,j+N/2^i}^{(t)} = \alpha_{i,j+N/2^i}^{(t)} \cdot g\left(\mathbf{R}_{i,j}^{(t)}, \mathbf{L}_{i+1,j}^{(t-1)}\right) + \mathbf{L}_{i+1,j+N/2^i}^{(t-1)}, \\ \mathbf{R}_{i+1,j}^{(t)} = \beta_{i+1,j}^{(t)} \cdot g\left(\mathbf{R}_{i,j}^{(t)}, \mathbf{L}_{i+1,j+N/2^i}^{(t-1)} + \mathbf{R}_{i,j+N/2^i}^{(t)}\right), \\ \mathbf{R}_{i+1,j+N/2^i}^{(t)} = \beta_{i+1,j+N/2^i}^{(t)} \cdot g\left(\mathbf{R}_{i,j}^{(t)}, \mathbf{L}_{i+1,j}^{(t-1)}\right) + \mathbf{R}_{i,j+N/2^i}^{(t)}, \end{cases}$$

(20)

where $\alpha_{i,j}^{(t)}$ and $\beta_{i,j}^{(t)}$ denote the right-to-left and left-to-right trainable scaling parameters, respectively.

For the training of scaling parameters, the commonly used loss function for supervised binary classification is binary cross-entropy (BCE) as shown in Fig. 4(a), which requires the information messages $\mathbf{u}$ for calculation as follows:

$$\mathcal{L}_{\text{BCE}}(\mathbf{u}, \mathbf{s}) = \frac{1}{N} \sum_{i=0}^{N-1} \left[ u_i \log \sigma(-s_i) + (1 - u_i) \right. \\ \left. \times \log(1 - \sigma(-s_i)) \right],$$

(21)

where $\sigma(z) = 1/(1 + e^{-z})$ is the sigmoid function. Besides, BCE is also adopted in [30] for the supervised learning of CRC-aided NN-BP decoder.

Then, based on the proposed syndrome loss, the unsupervised learning for neural network-based polar decoder with two different syndrome losses is shown in Fig. 4(b) and provided in Algorithm 1. Firstly, the randomly generated message $\mathbf{u}$ can be encoded by (5) and then the received signal $\mathbf{y}$ is obtained by transmitting the codeword over the channel with added noise. The log-likelihood ratios (LLRs), as the input for the BP decoder, can be transformed from $\mathbf{y}$ and given by:

$$\mathbf{llr} = 2\mathbf{y}/\sigma^2.$$

(22)

---

**Algorithm 1**: Unsupervised Learning for Neural Network-Based Belief Propagation Polar Decoder

**Input:** $\mathbf{llr}$, $A$, $T$, $\eta$
**Output:** $\boldsymbol{\alpha}$, $\boldsymbol{\beta}$
1: $\{\boldsymbol{\alpha}, \boldsymbol{\beta}\} \leftarrow \mathbf{1}$
2: **while** training stop criterion not met **do**
3:     $\mathbf{L}, \mathbf{R} \leftarrow$ initialize the NN-BP decoder($\mathbf{llr}$, $A$)
4:     $\mathbf{L}, \mathbf{R} \leftarrow$ NN-BP decoder($\mathbf{L}, \mathbf{R}, \boldsymbol{\alpha}, \boldsymbol{\beta}, T$)

**Algorithm 1.1: Proposed Frozen-Bit Syndrome Loss**
5:     **for** $t = 1$ **to** $T$ **do**
6:       **for** $j = 0$ **to** $N - 1$ **do**
7:         $s_{\text{froz},j}^t = \mathbf{L}_{n,j}^{(t)} + \mathbf{R}_{n,j}^{(t)}$
8:     $\{\boldsymbol{\alpha}, \boldsymbol{\beta}\} \leftarrow \text{SGD}\left(\boldsymbol{\alpha}, \boldsymbol{\beta}, \mathcal{L}_{\text{froz\_synd}}\left(\mathbf{s}_{\text{froz}}, \mathbf{H}_{\text{froz}}\right), \eta\right)$

**Algorithm 1.2: Proposed CRC-Enabled Syndrome Loss**
5:     **for** $t = 1$ **to** $T$ **do**
6:       **for** $j = 0$ **to** $N - 1$ **do**
7:         **if** $j \in A$ **do**
8:           $s_{\text{CRC},j}^t = \mathbf{L}_{0,j}^{(t)} + \mathbf{R}_{0,j}^{(t)}$
9:     $\{\boldsymbol{\alpha}, \boldsymbol{\beta}\} \leftarrow \text{SGD}\left(\boldsymbol{\alpha}, \boldsymbol{\beta}, \mathcal{L}_{\text{CRC\_synd}}\left(\mathbf{s}_{\text{CRC}}, \mathbf{H}_{\text{CRC}}\right), \eta\right)$

---

For each iteration of training, mini-batches of size $M$ of LLRs are operated in parallel to improve the convergence speed and ensure the stability of the training process. After $T$ iterations of BP decoding, the soft output from decoder can be obtained from updated messages $\mathbf{L}$ and $\mathbf{R}$ according to (16) and (18) for the calculation of frozen-bit syndrome loss and CRC-enabled syndrome loss, respectively. Then, the trainable scaling parameters $\boldsymbol{\alpha}$ and $\boldsymbol{\beta}$ are optimized through gradient descent on the used syndrome loss. In this work, the adopted algorithm for optimization is stochastic gradient descent (SGD), which iteratively updates the trainable parameters based on the gradient of the loss function as follows:

$$\boldsymbol{\theta}^{(j+1)} = \boldsymbol{\theta}^{(j)} - \eta \nabla_{\boldsymbol{\theta}} \mathcal{L}_{\text{synd}}\left(\boldsymbol{\theta}^{(j)}, \mathbf{s}, \mathbf{H}\right), \boldsymbol{\theta} = \{\boldsymbol{\alpha}, \boldsymbol{\beta}\}, \quad (23)$$

where $\boldsymbol{\theta}$ is the set of trainable parameters, $\eta > 0$ is the learning rate, and $\nabla_{\boldsymbol{\theta}} \mathcal{L}_{\text{synd}}$ is the gradient of the proposed syndrome loss function. After reaching a fixed number of training iterations or meeting some stop criterion, the well-trained parameters can be obtained, which achieves unsupervised learning without the knowledge of labeled data.

### B. Simulation Results

We first evaluate the capability of proposed syndrome loss on the proposed application scenario in Section IV.A. The experiments are implemented based on the Tensorflow framework which provides automatic differentiation of the loss function. Each simulation result is obtained with at least 1000 error blocks to make sure that the results are accurate and stable enough. For the following experiments, if there is no specific statement, all parameters and environment settings are based on Table II. Besides, we also adopt our previous work of recurrent neural network-based belief propagation (RNN-BP) polar decoder, which only requires 5 iterations for convergence [5]. The proposed recurrent architecture can force the network to learn shared weights among different iterations, which is helpful for reducing memory overhead and hardware complexity.

TABLE II
SIMULATION SETTINGS

| Parameters | Notation | Values |
|---|---|---|
| Encoding | $N, K, C$ | Polar Code (64,26,6) |
| Number of BP Iterations | $T$ | 5 |
| Modulation Category | - | BPSK |
| CRC Generator Polynomial | - | $x^6 + x^5 + 1$ |
| Training Codeword/SNR | - | 75,000 |
| Testing Codeword/SNR | - | 180,000 |
| Mini-batch Size | $M$ | 3,600 |
| Learning Rate | $\eta$ | 0.03 |
| Validation Ratio | - | 0.2 |
| Optimizer | - | Stochastic Gradient Descent |
| Training and Testing Environment | - | DL library of TensorFlow with NVIDIA GTX 1080 Ti GPU |

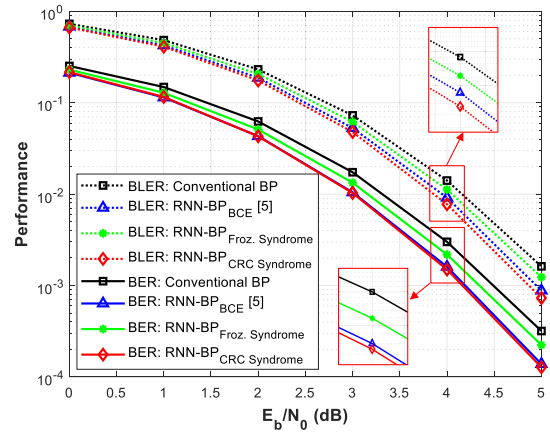

Fig. 5. Comparison of BER and BLER performance between the proposed two kinds of syndrome loss and binary cross-entropy (BCE).

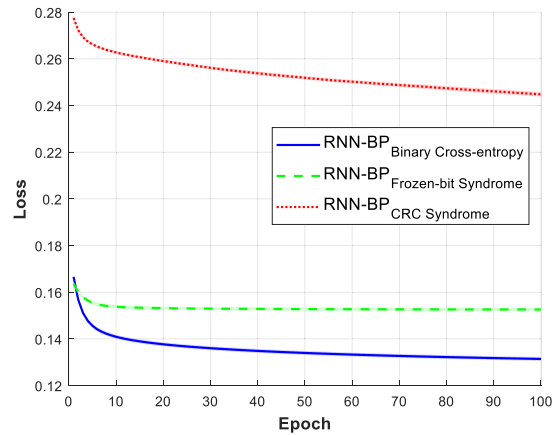

Fig. 6. Evolution of the validated loss between the proposed two kinds of syndrome losses and binary cross-entropy.

Furthermore, we adopt a shorter CRC for evaluation due to using a code length $N$ of 64. The utilized 6-bit CRC generator polynomial is $x^6 + x^5 + 1$ as reported in [17], which is also summarized in Table II. The $K$-bit information will first pass through the CRC system. Then, 6-bit check values, based on the remainder of the polynomial division of the information bits, will be attached for later check. Therefore, the total number of information bits is increased to $K + 6$. For simplicity, the CRC feature is only utilized to derive CRC-enabled syndrome loss for unsupervised learning without applying CRC-aided BP decoder in [30] to improve decoding performance.

*1) Performance for RNN-Based Polar Decoder:* In this experiment, we compare the bit error rate (BER) and block error rate (BLER) between the widely adopted supervised learning and unsupervised learning, which use binary cross-entropy (BCE) and the two kinds of proposed syndrome loss, respectively. Besides, the performance of conventional BP is also measured as a baseline reference to evaluate whether the different loss functions are all effective for the training of RNN-BP to improve the convergence speed. Thus, for a fair comparison, the number of BP iterations for both conventional BP and RNN-BP is set to 5 equally.

From Fig. 5, we can observe that the proposed CRC-enabled syndrome loss has a slightly better performance than BCE in terms of BER and has a bigger performance gap under BLER. Surprisingly, the performance of unsupervised learning is even better than supervised learning, which means that the decoder can merely learn from the code structure without the knowledge of transmitted codeword $c$. We surmise that the restrictions, added on the part of CRC factor graph, force the decoder to learn the decoding algorithm from the block level rather than the bit level. Therefore, CRC-enabled syndrome loss is more suitable for training than BCE and thus achieve better performance.

On the other hand, the performance of the proposed frozen-bit syndrome loss only has a slight improvement over conventional BP and is worse than supervised learning. This result may come from the fact that only the part of the frozen bits is constrained, which is not comprehensive enough for training the whole polar factor graph. However, it still proves that the proposed frozen-bit syndrome loss can exploit the nature of polar codes and also achieve unsupervised learning when the transmitted codeword is not provided.

*2) Convergence Speed for RNN-Based Polar Decoder:* In this experiment, we evaluate stability and convergence speed by comparing the evolutions of the validated losses and validated block error rate during the first 100 training epochs. Due to the unstable of validated losses and block error rate, both values are averaged over 30 seeds with signal-to-noise ratio (SNR) ranging from 0dB to 5dB. The shaded areas around the curves correspond to one standard deviation in each direction.

In general, we can consider that supervised learning has better convergence speed over unsupervised learning due to the demanded ground truth of message $u$ for training. However, from Fig. 6 and Fig. 7, we can find out that both methods have fast convergence speeds and significant improvement of BLER during the first 10 epochs with small variance, which demonstrate the feasibility and stability of the proposed syndrome loss for unsupervised learning. Besides, the slopes of loss and BLER based on CRC-enabled syndrome loss are greater than BCE, which demonstrates its great potential for further performance enhancement.

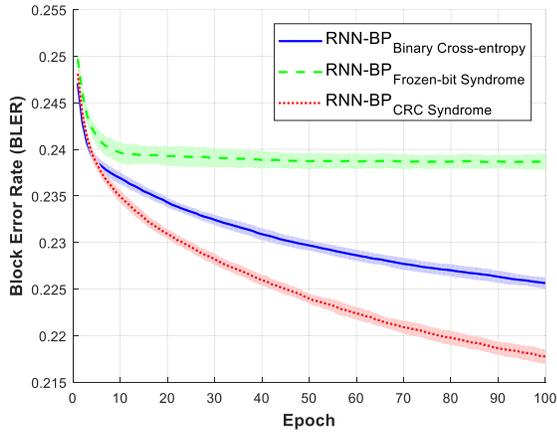

Fig. 7. Evolution of the validated BLER between the proposed two kinds of syndrome losses and binary cross-entropy.

## V. APPLICATION SCENARIO II: SYNDROME-ENABLED BLIND EQUALIZER WITH JOINT OPTIMIZATION MECHANISM

In Section IV.A, to reduce the variability, we consider an ideal channel with only additive white Gaussian noise. Therefore, the capability of proposed syndrome loss for unsupervised learning can be carefully analyzed and compared. However, for channel coding, we can easily obtain the labeled data by randomly generating message $u$ and transmitting over simulated AWGN channel. Therefore, unsupervised learning for neural network-based decoder is nice to have but not necessary. Therefore, after confirming the capability of proposed syndrome loss for training neural network-based decoder, we propose another application scenario to demonstrate the benefit and the prominent of syndrome loss.

### A. Proposed Syndrome-Enabled Unsupervised Learning for Blind Equalizer With Joint Optimization Mechanism

In this scenario, we consider the block fading channel, which means the channel impulse response $h$ is assumed to be constant through the whole block of multiple symbols but may vary from one block to the other [33]. Thus, conventional adaptive MMSE equalizer demands training sequence for channel adaptation to avoid significant degradation of performance. However, the transmission of training sequence will also occupy transmission resources and reduce spectral efficiency.

Therefore, based on the proposed syndrome loss, we propose a syndrome-enabled blind equalizer as shown in Fig. 1(c), which can finetune filter parameters $\hat{h}$ without transmission of training sequence. In this work, we dedicate to demonstrate the capability of proposed syndrome loss for blind equalization under block-fading channel. Therefore, a filter with convolutional operations is good enough for eliminating ISI channel effect and can be fairly compared with conventional MMSE equalizer [18]. Therefore, the only difference between MMSE equalizer and the proposed blind equalizer is that the updating of filter coefficients is different. The former depends on training sequence and attempts to minimize the error between equalized signal and training sequence based on the measure of mean square error (MSE). The latter takes advantage of domain knowledge in communication to penalize non-valid output from decoder based on the proposed syndrome loss, which can truly reflect the overall system performance.

Consequently, in addition to the benefits of improved spectral efficiency, the utilization of syndrome loss also achieves joint optimization of equalizer and decoder. As the concept of autoencoder for end-to-end learning systems has been proposed in [7]–[9], many researchers try to break the conventional rules of individual block design by jointly optimizing the whole communication systems, which achieves global optimum instead of local optimum. Similarly, the authors in [31]–[32] focus on the receiver and propose a neural network for joint channel equalization and decoding, which demonstrates that the network has the ability to simultaneously address various channel effects and learn the complicated decoder function with a better performance.

Although we do not adopt a neural network-based equalizer in this work, the proposed method can not only eliminate channel fading but also force the equalized signal to be more suited for the subsequent decoder, thereby further enhancing the overall system performance. For the future extension under more complicated channel conditions or more severe hardware impairments, the equalizer part can be easily replaced with a powerful neural network and still benefit from our proposed method.

The proposed syndrome-enabled blind equalizer with joint optimization mechanism is given in Algorithm 2 with two different syndrome losses. Firstly, the equalized signal $\hat{x}$ can be obtained by convolving the received signal $y$ with filter $\hat{h}$ and transformed to LLRs as input for NN-BP decoder. Note that the meaning of joint optimization is to update filter $\hat{h}$ to force the equalized signal more suited for the subsequent decoder. It does not mean to jointly optimize the parameters of filter and NN-BP decoder. Therefore, the well-trained parameters $\alpha$ and $\beta$ can be loaded into NN-BP decoder for better convergence speed and set as frozen during the updating of filter coefficients.

For each training iteration, a mini-batch of size $M$ of received signals, which corresponds to the number of available received signals under the same channel state, is utilized for training. Therefore, the more stable the channel, the more received signals can be used to improve the estimation accuracy and ensure the stability of training process. The filter coefficients are optimized through stochastic gradient descent on the proposed syndrome loss. After meeting some stop criterion, the filter coefficients are updated to eliminate current channel effect with improved quality

**Algorithm 2**: Syndrome-Enabled Blind Equalizer With Joint Optimization Mechanism

**Input:** $y$, $\alpha$, $\beta$, $A$, $T$, $\eta$
**Output:** $\hat{h}$
1: $\hat{h} \leftarrow$ initialize the filter coefficients
2: **while** training stop criterion not met **do**
3: $\quad \hat{x} = y \otimes \hat{h}$
4: $\quad llr = 2\hat{x}/\sigma^2$
5: $\quad \mathbf{L}, \mathbf{R} \leftarrow$ initialize the NN-BP decoder($llr$, $A$)
6: $\quad \mathbf{L}, \mathbf{R} \leftarrow$ NN-BP decoder($\mathbf{L}, \mathbf{R}, \alpha, \beta, T$)

**Algorithm 2.1: Proposed Frozen-Bit Syndrome Loss**
7: $\quad$ **for** $t = 1$ **to** $T$ **do**
8: $\quad\quad$ **for** $j = 0$ **to** $N - 1$ **do**
9: $\quad\quad\quad s^t_{\text{froz},j} = \mathbf{L}^{(t)}_{n,j} + \mathbf{R}^{(t)}_{n,j}$
10: $\quad \hat{h} \leftarrow \text{SGD}\left(\hat{h}, \mathcal{L}_{\text{froz\_synd}}(s_{\text{froz}}, \mathbf{H}_{\text{froz}}), \eta\right)$

**Algorithm 2.2: Proposed CRC-Enabled Syndrome Loss**
7: $\quad$ **for** $t = 1$ **to** $T$ **do**
8: $\quad\quad$ **for** $j = 0$ **to** $N - 1$ **do**
9: $\quad\quad\quad$ **if** $j \in A$ **do**
10: $\quad\quad\quad\quad s^t_{\text{CRC},j} = \mathbf{L}^{(t)}_{0,j} + \mathbf{R}^{(t)}_{0,j}$
11: $\quad \hat{h} \leftarrow \text{SGD}\left(\hat{h}, \mathcal{L}_{\text{CRC\_synd}}(s_{\text{CRC}}, \mathbf{H}_{\text{CRC}}), \eta\right)$

TABLE III
SIMULATION SETTINGS FOR FADING CHANNEL

| Parameters | Notation | Values |
|---|---|---|
| Number of Channel Taps | $L$ | 3 |
| Channel Distance Vector | $d$ | [1.5, 2.0, 2.7] |
| Path-loss Exponent | $\gamma$ | 2 |
| Channel Variance | $\sigma_h^2$ | 1 |
| Number of Filter Coefficients | $F$ | 5 |

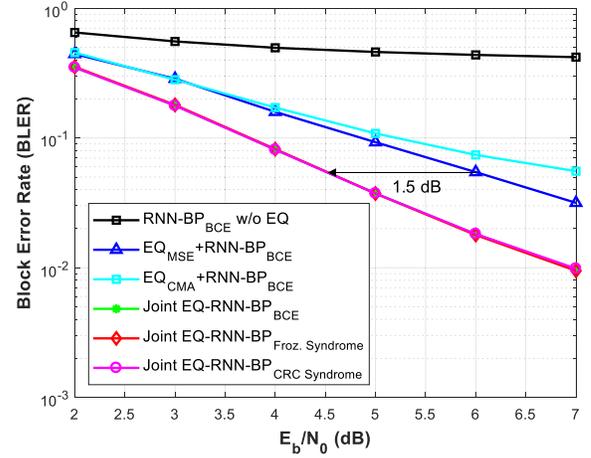

Fig. 8. Comparison of BLER performance between the proposed syndrome-enabled blind equalizer and prior works under time-invariant multipath channels.

of equalized signals. Therefore, our proposed approach can achieve blind equalization without the transmission of pilot signals.

### B. Simulation Results

Now, we evaluate the performance on the proposed application scenario in Section V.B. We evaluate and compare the proposed syndrome-enabled blind equalizer with the state-of-the-art adaptive MMSE equalizer [18] and the mechanism of online label recovery [13]. Besides, the performance of blind adaptive CMA is also compared, which is a standard and representative approach for blind channel equalization [21].

*1) Performance for Proposed Blind Equalizer Under Time-invariant Multipath Channel:* Although the blind equalizer can avoid the transmission of training sequence, it still relies on enough received signals under the same channel state, for accurate estimation of channel characteristic and stable updating of filter coefficients. Therefore, we first suppose that the channel is time-invariant, so we have enough stable signals for channel adaptation. Besides, to make the simulation results generalized enough, we average the evaluation metrics over 30 different channel conditions. For each channel, the channel impulse response is randomly generated by:

$$h_l = d_l^{-\gamma} \times r, l = 0, \ldots, L - 1, \quad (24)$$

where $\gamma$ denotes path-loss exponent, $r$ is randomly generated from normal distribution with zero mean and variance $\sigma_h^2$, and $d$ is the distance vector of the multipath from transmitter to receiver, which is highly correlated to the received signal strength [34]. Finally, the power of channel impulse response will be normalized to 1. The setups of channel parameters are shown in Table III.

Under time-invariant multipath channels, we have enough training sequences for accurate channel estimation. Therefore, the mechanism of online label recovery is skipped in this part. Besides, we also compare with the method without using equalizer as shown in Fig. 8. The subscript in each label represents the loss function used for optimization. It is worth to notice that the scaling parameters of RNN-BP is well-trained beforehand under AWGN channel and freezed in this system without any adjustment.

Firstly, from Fig. 8, we can find out that the performance of RNN-BP decoder without equalizer is significantly worse than that of RNN-BP decoder with MMSE equalizer, which indicates that RNN-BP decoder cannot effectively eliminate channel effect despite it has trainable scaling parameters.

Secondly, the performance of the proposed blind equalizer with joint optimization mechanism is evaluated under different loss functions, including binary cross-entropy and two kinds of proposed syndrome losses. From Fig. 8, we can observe that the performance is almost the same and therefore overlapped on the graph, which means that the loss can be effectively backpropagated to the part of filter for parameter adjustment.

Thirdly, we can observe that the blind CMA equalizer, with sufficient received signal for accurate channel adaptation, is only slightly worse than MMSE. However, the most impressive finding is that the performance of the proposed method has about 1.5 dB gain over MMSE equalizer with RNN-BP decoder. We want to emphasize again that the only difference between these two methods is the optimized loss function. One is based on the mean square error between training sequence and equalized signal, while the other is the proposed syndrome loss calculated from the decoded output, which can truly reflect the overall system performance. The cascaded RNN-BP decoder has exactly the same architecture and parameters. Therefore, different loss functions result in different filter coefficients and thus different performance results. The simulation results demonstrate the great benefit and potential for joint optimization mechanism to outperform conventional block based design. In the following experiments, we will further confirm the conclusion.

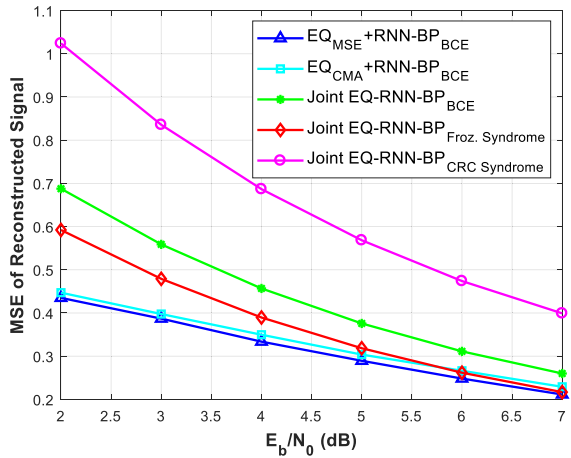

Fig. 9. Comparison of reconstructed signal quality between the proposed syndrome-enabled blind equalizer and prior works under time-invariant multipath channels.

*2) Quality of Reconstructed Signal:* To further analyze what happened for the mechanism of joint optimization, we utilize the measurement of mean square error (MSE) to evaluate the quality of reconstructed signal after equalization. The metric can be defined as:

$$\mathcal{L}_{\text{MSE}}(\boldsymbol{x}, \hat{\boldsymbol{x}}) = \frac{1}{N} \sum_{i=0}^{N-1} (x_i - \hat{x}_i)^2. \quad (25)$$

From Fig. 9, as expected, the MMSE equalizer has the best reconstructed signal quality because the utilized LMS algorithm is dedicated to minimizing MSE. Therefore, it directly reflects on the quality of the reconstructed signal. For the CMA equalizer, it has slightly worse MSE due to the unavailable training sequence.

On the other hand, although the three different kinds of loss functions for joint optimization mechanism have the same performance as shown in Fig. 8, the quality of reconstructed signals has an obvious difference. This means that the different loss functions have their own optimization curves for filter, resulting in different qualities of reconstruction, but all of them can achieve almost the same BLER by directly optimizing the decoder output.

These key findings can be summarized and concluded as below. In the past, we may assume that the best MSE of the reconstructed signal can result in the best signal for the decoder, thus achieving the best system performance. However, from the simulation results, we demonstrate that the best MSE of the reconstructed signal may not be most suitable for the decoder's operations and cannot truly reflect the decoding performance. Instead, by optimizing from the system level with a "joint optimization" mechanism, the coefficients of the equalizer filter can be updated in an optimal manner to minimize the loss of the decoder output. Thus, the equalized signal can provide more insightful information for the channel decoder to improve the overall system performance.

*3) Performance for Proposed Blind Equalizer Under Block Fading Channel:* After demonstrating the great performance of proposed syndrome-enabled blind equalizer under time-invariant multipath channel, we start to evaluate the performance under block fading channel. In our case, the block means a codeword with $N$ symbols. Thus, the fading process is constant over the block of $N$ symbols and it is statistically independent between the blocks, which is more consistent and suitable for real communication systems with slowly moving [33]. It means that the number of received signal blocks under the same channel condition is limited. In this part, we include the mechanism of online label recovery to obtain labeled data for the updating of MMSE equalizer [13]. Besides, due to the similar performance between binary cross-entropy and syndrome loss in Fig. 8 and the unrealistic requirement of messages $\boldsymbol{u}$ for training in communication systems, we exclude this method from the graphs. In addition, the method of MMSE equalizer with training sequences can observe all of the transmitted codeword $\boldsymbol{x}$, which is also unrealistic but can be provided as the ideal case and the best performance that MMSE equalizer can achieve. In Fig. 10, we average the BER over 100 different channel conditions. To simulate the block fading channel, we constrain the number of received signal block $M$ for each channel condition. The bigger $M$ means the channel is slow fading and thus we can obtain more received signal blocks for channel adaptation. On the contrary, the smaller $M$ means the channel is fast fading and thus the available signal block for each channel condition is less.

From Fig. 10, we can observe that MMSE equalizer without training sequence has the worst performance due to the lack of training sequences for channel adaptation. However, $M = 100$ is enough for the proposed blind equalizer to outperform the MMSE equalizer with training sequence by about 1.3 dB as shown in Fig. 10(a), which benefits from the mechanism of joint optimization. Besides, the performance of CRC-enabled syndrome loss is better than frozen-bit syndrome loss, which means that the CRC-enabled syndrome loss can update the filter coefficients more efficiently under limited received signals.

Besides, there is a huge gap between ideal MMSE equalizer and MMSE equalizer with online label recovery. The gap comes from the incorrectly decoded output results in codewords that are incorrectly re-encoded, and thus even deteriorates the filter coefficients. On the other hand, the representative blind CMA equalizer has slightly better performance than the mechanism of online label recovery. However, $M = 100$ is still not enough for accurate channel adaptation and thus also has a huge degradation compared to adaptive MMSE equalizer.

Furthermore, as the channel varies faster, the deterioration becomes more obvious and severe as shown in Fig. 10(b) and Fig. 10(c). The performance of both online label recovery and CMA is even worse than MMSE without training sequence when $M = 1$ due to the lack of training sequence and the unstable training process. However, we can still observe that there is about a 1 dB and 2 dB performance gain of proposed blind equalizer compared to the mechanism of online label recovery under $M = 10$ and $M = 1$, respectively, which demonstrates the applicability and reliability of the proposed syndrome-enabled blind equalizer.

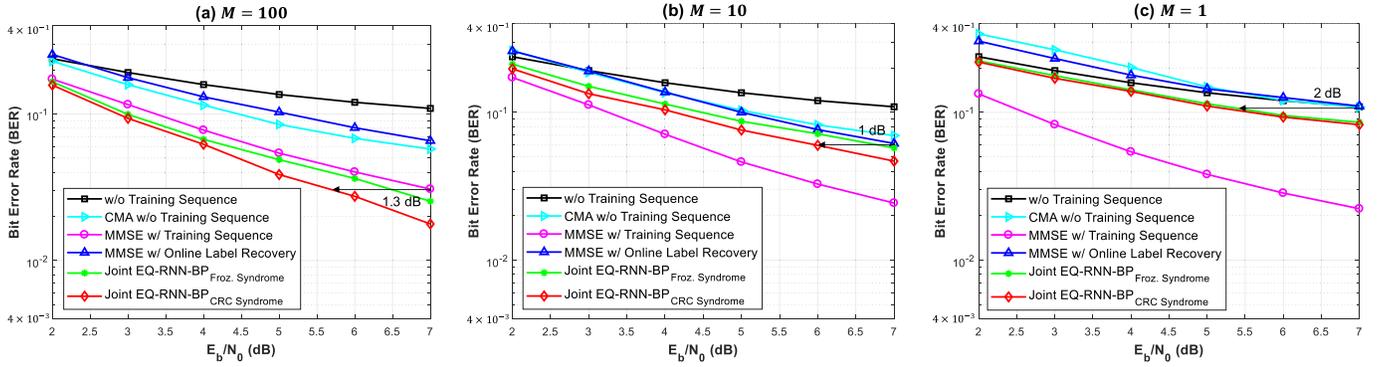

Fig. 10. Comparison of BER performance between the proposed syndrome-enabled blind equalizer and prior works under the different number of training/received signals: (a) $M = 100$, (b) $M = 10$, and (c) $M = 1$.

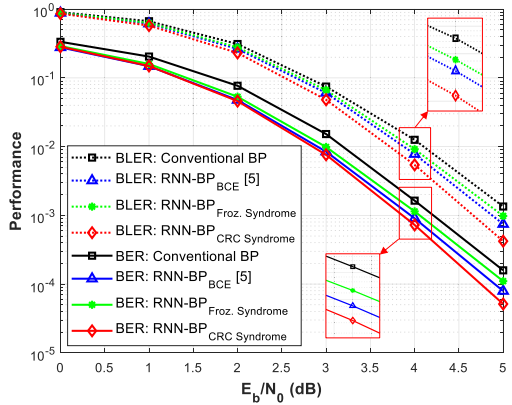

Fig. A.1. Comparison of BER and BLER performance between the proposed two kinds of syndrome losses and binary cross-entropy (BCE).

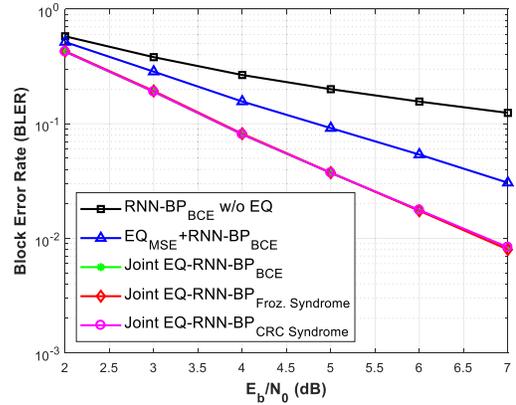

Fig. A.2. Comparison of BLER performance between the proposed syndrome-enabled blind equalizer and prior works under time-invariant multipath channels.

## VI. CONCLUSION

In this paper, we propose two kinds of modified syndrome losses, which make unsupervised learning possible for systems with polar codes. We propose two application scenarios to evaluate the capability of the proposed syndrome losses. In the case of training neural network-based polar decoder, the proposed CRC-enabled syndrome loss can even outperform prior works based on supervised learning. In the second case, the proposed syndrome-enabled blind equalizer can avoid the transmission of training sequences under block fading channel and achieve global optimum via joint optimization mechanism. Both cases show that the domain-specific syndrome loss provides a new tool for the paradigm of machine learning-assisted communication systems to overcome many realistic problems, such as channel variations, without incremental transmission overhead.

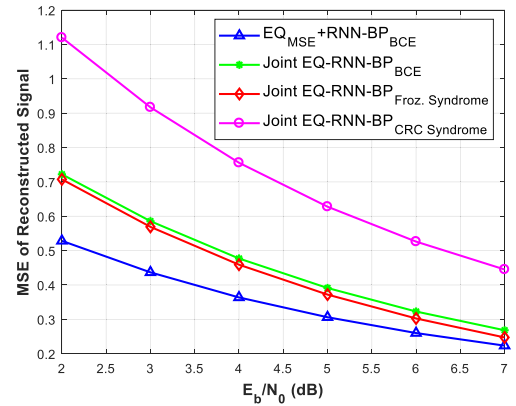

Fig. A.3. Comparison of reconstructed signal quality between the proposed syndrome-enabled blind equalizer and prior works under time-invariant multipath channels.

## APPENDIX

To demonstrate the scalability of our proposed approach, we also repeat the simulations by increasing code length from 64 to 128 in Fig. A1 to Fig. A3. We can observe that the simulation results under $N = 128$ have a similar trend as $N = 64$. That is, the proposed frozen-bit syndrome loss and CRC-enabled syndrome loss are independent with code length and can be easily extended to longer code length.

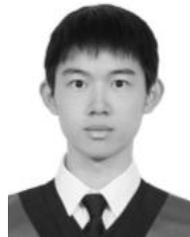

**Chieh-Fang Teng** (Student Member, IEEE) received the B.S. degree in electrical engineering from National Taiwan University, Taipei, Taiwan, in 2017, where he is currently pursuing the Ph.D. degree with the Graduate Institute of Electronics Engineering. His research interests are in the areas of the Internet of Things (IoT), VLSI architecture for DSP, and machine learning assisted wireless communication systems design.

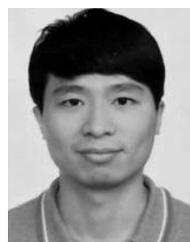

**Yen-Liang Chen** received the B.S. degree in communication engineering from National Chiao Tung University, Hsinchu, Taiwan, in 2005, and the Ph.D. degree in electronic engineering from National Taiwan University, Taipei, Taiwan, in 2011.

He is currently a Technical Manager with MediaTek Inc., Taiwan. His major research interests include digital communication system designs, digitally assisted RF design techniques, and machine-learning assisted communication system designs.